# Step Stone Effect: An *sp* anti-bonding state Mediated Long-Range Ferromagnetism in Cr-doped Carrier-Free $Bi_2Te_3$

Chunkai Chan, Xiaodong Zhang, Yiou Zhang, Kinfai Tse, Bei Deng, Jingzhao Zhang, and Junyi Zhu[*]

Department of Physics, Chinese University of Hong Kong, Hong Kong SAR, China

Despite the recent success in the realization of the quantum anomalous Hall effect, the underlying physical mechanism of the long-range ferromagnetism is still unclear. Based on our density functional theory calculations, we discovered an intriguing long-range ferromagnetic order in Cr-doped, carrier-free $Bi_2Te_3$, with the separation between dopants more than 8 Å. We found that this magnetic coupling is facilitated by an anti-bonding state originated from the Te 5*p* and Bi 6*s* state, despite this state lies below the valence band maximum. We name such a state as a step stone state. An effective electron hopping model is proposed to explain this mechanism. This novel mechanism sheds light on the understanding of long-range ferromagnetism in insulators and may lead to the realization of the long-range magnetic order in dilute magnetic semiconductors.

The carrier-independent ferromagnetism (FM) and anti-ferromagnetism (AFM) are of great technological significance in applications of dilute magnetic semiconductors (DMS). In such applications, the ideal purpose of free carrier manipulation is only to realize the desired electronic functions with magnetic properties unchanged. However, realistically, it's almost unavoidable to tune both properties, until very recently, it has been demonstrated that a long-range FM independent of both polarity and density of free carriers and even without carriers can exist in Cr-doped $(Sb,Bi)_2Te_3$ thin film, leading to the discovery of quantum anomalous Hall effects [1-10]. The key to the success is due to the long range FM, however, the underlying physical mechanism of such long-range interaction is still largely unclear, except that the enhanced van Vleck paramagnetism due to the inverted band structures was proposed [11]. Such claim was based on the calculations of $Bi_2Se_3$ with a small supercell of 60 atoms, where the magnetic dopant separations cannot be considered as long range [11]. More importantly, Cr-doped $Bi_2Se_3$ [12-13] fails to observe any significant long-range ferromagnetic order.

$(Sb,Bi)_2Te_3$ and $Bi_2Se_3$ are both topological insulators with conducting surface states and insulating bulk states, protected by time reversal symmetry [14]. The introduction of magnetic dopants, such as Cr, may destroy the time reversal symmetry and the surface states [4, 7, 11, 15-26]. No any theoretical analysis has been performed to study the difference between these two similar material systems.

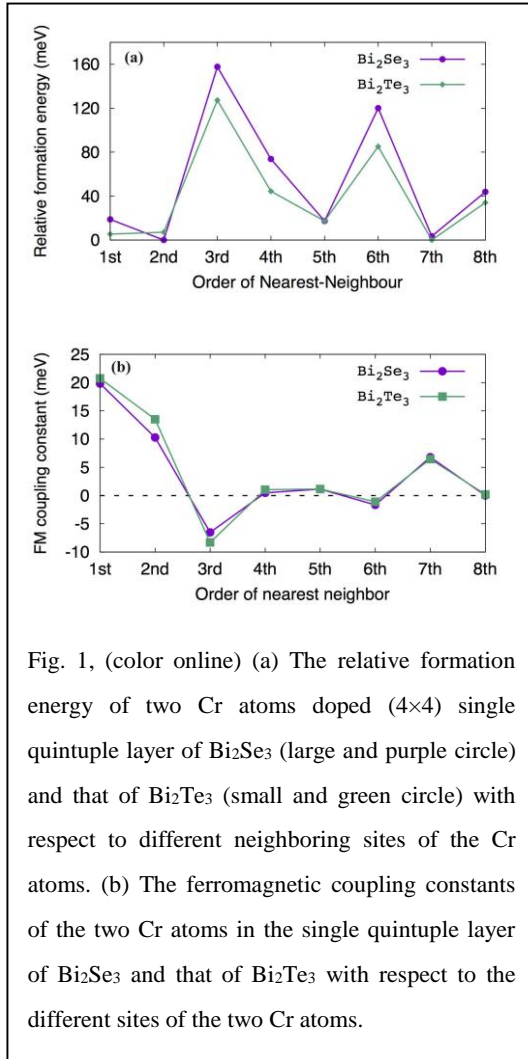

Fig. 1, (color online) (a) The relative formation energy of two Cr atoms doped (4×4) single quintuple layer of $Bi_2Se_3$ (large and purple circle) and that of $Bi_2Te_3$ (small and green circle) with respect to different neighboring sites of the Cr atoms. (b) The ferromagnetic coupling constants of the two Cr atoms in the single quintuple layer of $Bi_2Se_3$ and that of $Bi_2Te_3$ with respect to the different sites of the two Cr atoms.

In this letter, to address the above questions, we performed *ab-initio* calculations, based on density functional theory (DFT) implemented in VASP code with a sufficiently large simulation cell [27] [see computational details in Supplementary Information (SI)]. To avoid further complexity of alloy effects in $(Sb,Bi)_2Te_3$, we used $Bi_2Te_3$ as a model system and compared the magnetic order of Cr doped $Bi_2Te_3$ and $Bi_2Se_3$. Later, we also incorporated a few Sb atoms as dopants into $Bi_2Te_3$ to study the effect of Sb. To our surprise, we discovered a novel long-range ferromagnetic order in carrier free $Bi_2Te_3$, which is the most stable configuration. This finding is independent upon the band topology. The similar ferromagnetic order is relatively unstable for $Bi_2Se_3$, consistent with the experimental results [13]. Then, we proposed a mechanism to qualitatively explain the intriguing interaction based on an extended Hubbard model [28].

To study the long-range magnetic order of the Cr-doped $Bi_2Se_3$ and $Bi_2Te_3$ thin films, we first replaced two Bi atoms with two Cr atoms in the 4×4 host cell that contains one quintuple layer. We performed both GGA and GGA+U with spin orbit coupling taken into account (see SI for details), both methods yield no significant difference in formation energies and magnetic coupling strength (See SI). Therefore, throughout this main text, we show only the GGA results. The relative formation energy results are listed in Fig. 1.

The relative formation energy is defined as the energy difference between the formation energy of each configuration and that of the most stable configuration. As functions of the distance between the neighboring Cr atoms, the relative formation energy of $Cr_{Bi}$ in $Bi_2Se_3$ and that in $Bi_2Te_3$ show generally similar trends. Significant formation energy drops on the seventh neighboring sites were found in both systems. This configuration is the global minimum in $Bi_2Te_3$, which is even more stable than the first nearest neighbor one. Later, we also calculated the ferromagnetic coupling constants in respect to the different neighboring sites, as shown in Fig. 1 (b). The coupling constants are defined as half of the energy difference between the two Cr atoms with the same spins (FM) and the two Cr atoms with opposite spins (AFM). We found that the seventh nearest neighbor configurations favor FM and yield coupling constants about 6 meV for both systems, despite the fact that the coupling strength is lower than the first or second nearest neighboring configurations. These findings are surprising because in common DMS, the coupling strength usually decays fast in respect to the separation distances [29]. In our seventh nearest neighbor configuration, the two Cr atoms are separated by two anion atoms and one Bi atom.

Since the global minimum of the formation energy of Cr atoms in the $Bi_2Se_3$ thin film is the second nearest neighboring configuration, Cr atoms tend to form clusters in $Bi_2Se_3$, leading to nonmagnetic or paramagnetic orders. On the contrary, the global minimum of the formation energies of Cr atoms in the telluride suggests a long range FM order. These results qualitatively agree with the experimental observations [2, 12-13,30-31].

To understand the origin of such long-range FM, we further studied the intrinsic electronic properties of $Bi_2Te_3$. The projected density of states (pDOS) of Bi 6$s$ and Te 5$p$ are shown in Fig. 2 (a). Despite the local symmetry difference, two Te atoms at different sites yield similar 5$p$ orbital components, so we choose one Te atom to show these orbitals. From the pDOS figure, it's clear that the dominant component of the valence band is Te 5$p$. Note that, however, a small component of Bi 6$s$ state is found to be below the valence band maximum (VBM), although the major component of the Bi 6$s$ orbital is deep under the VBM. Despite the fact that the Bi 6$s$ orbital is mainly a lone pair state, a small component of it couples with the Te 5$p$ orbital and forms an occupied $sp$ anti-bonding $(anti-sp\sigma)$ state.

Next, we studied the single Cr-doped $Bi_2Te_3$. Cr dopant has 3 unpaired 3$d$ electrons with a magnetic moment of 3. The pDOS of the Cr 3$d$ orbitals and the 6$s$ orbital of the Bi next to the Cr dopant were calculated, as shown in Fig. 2 (b). The significant asymmetric nature of the spin up and spin down portion of the Bi 6$s$ orbital suggests that it is magnetized by the nearby Cr atom. Therefore, the $anti-sp\sigma$ state is also magnetized. It should be noted that the $anti-sp\sigma$ state is an intrinsic property of $Bi_2Te_3$ (see SI for the details of the partial charge of that state), and the role of Cr dopant is to magnetize it, but not to create it. The pDOS of Cr 3$d$ also indicates that there is a small occupation of spin down component, leading to the magnetization the surrounding atoms.

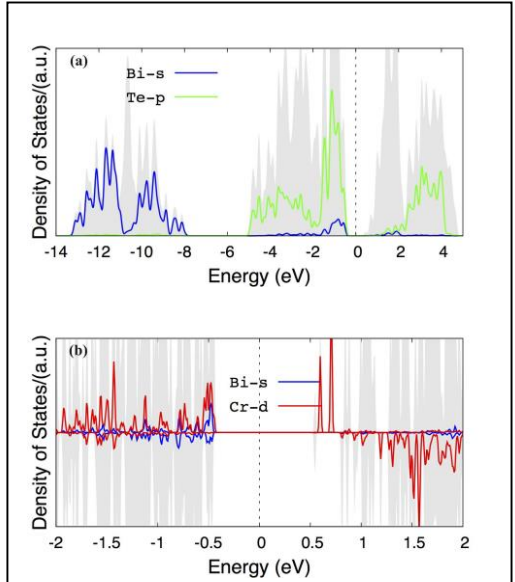

Figure 2. (Color online) (a) The pDOS of the Bi 6$s$ orbital (blue) and Te 5$p$ orbital (green) in pure $Bi_2Te_3$. The total density of states is shown in the grey shading background. And the dotted line shows the Fermi level. (b) The pDOS of the Cr 3$d$ orbital (red) and the 6$s$ orbital of Bi (blue) (Bi atom on the Cr-Te-Bi-Te-Cr path) in single Cr-doped $Bi_2Te_3$. The grey background is the total DOS of single Cr-doped $Bi_2Te_3$, and the positive and negative values represent the pDOS of spin up and down components, respectively.

To understand the magnetization of surrounding atoms, we further calculated the spin density of single Cr-doped $Bi_2Te_3$, as shown in Fig. 3 (a), where only the spin polarized atoms are displayed. We found that the neighboring $p$ orbitals of Te are magnetized along the bonding direction between Te and Cr, which we labele as $p_z$ orbital, note that z axis is just along the local bonding direction. The total magnetization of all the Te $p$ orbitals is anti-parallel to the Cr magnetic moment. These results are in good agreement with the experimental observations [32]. The spin density also suggests that the spin up electron cloud near the Bi atom is the $anti-sp\sigma$ state, which also couples to the Cr atom. Interestingly, the Cr-Te-Bi chain forms a right angle, as shown in Fig. 3 (b). Note that such chains are fundamental building blocks to form the long-range magnetic order, as further studies suggest.

The spin polarized results can be qualitatively understood based on an electron hopping mechanism. Since the electronic enviroment around the Cr atom is approximately $O_h$ symmetry, the Cr 3$d$ orbitals are splitted into filled spin up $t_{2g}$ states and empty $e_g$ states. Due to the local symmetry, the Te $p_z$ orbital is only coupled to the Cr $e_g$ states. Since the energy of the spin up $e_g$ states is lower than that of the spin down component, the spin up electron of the Te $p_z$ orbital hops to the $e_g$ state. As a result, the $p_z$ orbital is magnetized towards spin down. Since the $t_{2g}$ states are already occupied by one spin up electron, only spin down electron in the $anti-sp\sigma$ state hops to the $t_{2g}$ states. As a result, the $anti-sp\sigma$ state is magnetized towards spin up.

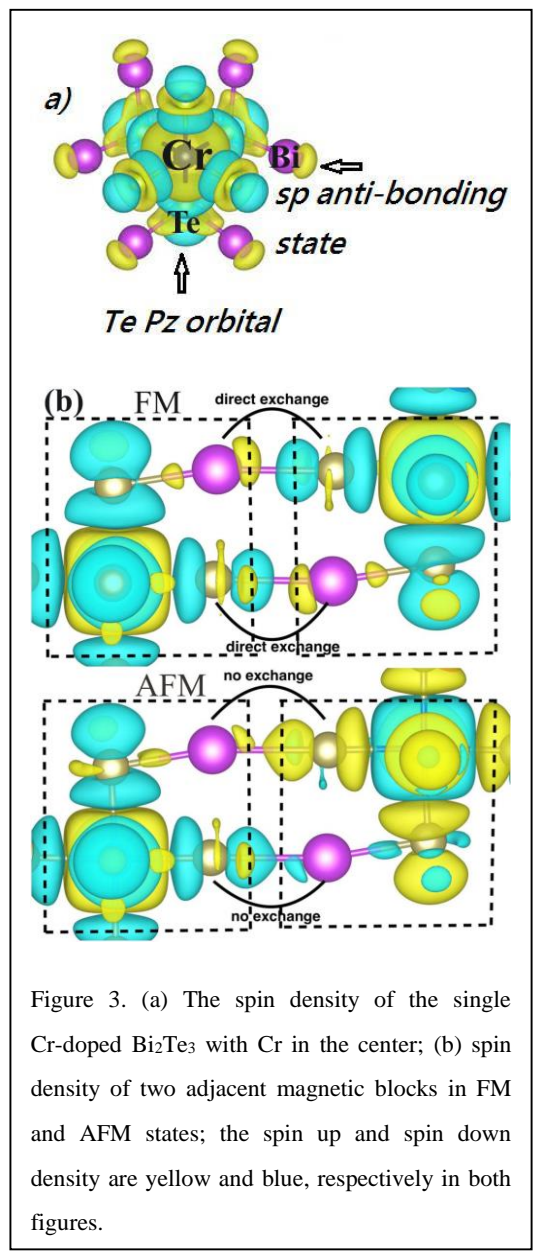

Figure 3. (a) The spin density of the single Cr-doped Bi$_2$Te$_3$ with Cr in the center; (b) spin density of two adjacent magnetic blocks in FM and AFM states; the spin up and spin down density are yellow and blue, respectively in both figures.

These results indicate that the magnetic moments of the Cr atom are not fully localized at the Cr site and a small portion of them are distributed in the surrounding orbitals. Still, the magnetic moment in the Te $p$ states cannot be interpreted as spin polarized hole state [32], because the Cr-doped Bi$_2$Te$_3$ and Bi$_2$Se$_3$ thin films remain carrier-free.

As shown in Fig. 3 (b), when two of these building blocks are placed in adjacent to each other, long-range ferromagnetic interaction can occur, with the two Cr atoms at seventh nearest neighboring sites. In the ferromagnetic configuration, the spin of the $anti-sp\sigma$ orbitals in one block and the adjacent $p_z$ orbitals in the other block are anti-parallel. Therefore, electron hopping between these orbitals will lower their kinetic energy. However, in the anti-ferromagnetic configuration, these two spins are parallel and no electron hopping occur to lower the energy. As a result, the ferromagnetic configuration is energetically favorable. Such building blocks further connect and build up a strong ferromagnetic order in the whole crystal. **In this novel scheme, the anti-bonding state near the middle cation serves as an important step stone to**

**mediate the magnetic interaction.** Therefore, we name this effect as a step stone effect and these Bi atoms connecting the two building blocks are at step stone sites.

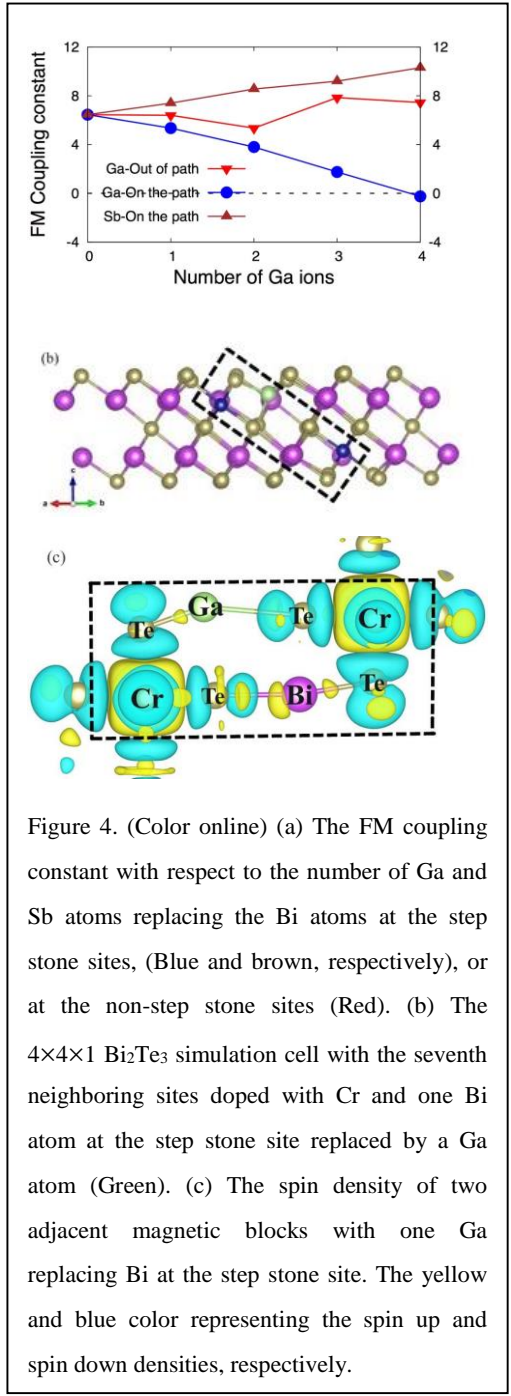

Figure 4. (Color online) (a) The FM coupling constant with respect to the number of Ga and Sb atoms replacing the Bi atoms at the step stone sites, (Blue and brown, respectively), or at the non-step stone sites (Red). (b) The 4×4×1 $Bi_2Te_3$ simulation cell with the seventh neighboring sites doped with Cr and one Bi atom at the step stone site replaced by a Ga atom (Green). (c) The spin density of two adjacent magnetic blocks with one Ga replacing Bi at the step stone site. The yellow and blue color representing the spin up and spin down densities, respectively.

To further validate our new mechanism, we replaced Bi atoms at the step stone sites with Ga atoms that do not have lone pair electrons, but share the same valence as Bi atoms. Due to periodic boundary condition, the two adjacent building blocks in 4×4×1 $Bi_2Te_3$ supercell have four step stone sites. We replaced one, two, three, and all of the Bi atoms at the sties with Ga atoms and calculated the magnetic coupling strength, as shown in Figure 4 (a). The magnetic coupling constant decrease approximately linearly and finally vanish when all four Bi atoms are replaced. In contrary to the Ga replaced case, the magnetic interaction remains approximately unchanged when Sb substitutes the step stone Bi, which is also qualitatively consistent with the FM order observed in $(Sb,Bi)_2Te_3$. To further demonstrate the unique property of the Bi atoms at the step stone sites, we also replaced the Bi atoms that are not at the step stone sites, but still in the same quintuple layer, with Ga atoms and found that ferromagnetic order preserves since the coupling constants almost do not change even when four non-step-stone Bi atoms are replaced. These results confirm our model.

Finally, to visualize the disappearance of the $anti-sp\sigma$ state when Bi at the step stone site is replaced, we calculated the spin density of two touching magnetic blocks with one Bi replaced by a Ga atom, as shown in Fig. 4 (c). It clearly shows that the Ga atom does not have a spin polarized $anti-sp\sigma$ orbital, which still can be seen around the remaining Bi atom. These results also suggest that the missing of magnetized states around the cation atoms may be the fundament reason to prevent long range magnetism in carrier free DMS.

To conclude, we have found a stable carrier independent long-range ferromagnetic interaction in $Bi_2Te_3$, which agrees well with experimental results [2,30,31]. An electron hopping mechanism based on the intrinsic $anti-sp\sigma$ state was proposed to qualitatively explain this intriguing finding. We expect this mechanism to be general in compounds with the proper anti-bonding states due to the cation lone pair and anion *p* orbital. This discovery may greatly enhance the understanding of the hidden and carrier free long-range magnetism, which we believe is critical in the developments of DMS and spintronic devices.

We are grateful for the financial support of Chinese University of Hong Kong (CUHK) (Grant No.4053084), University Grants Committee of Hong Kong (Grant No. 24300814), and start-up funding of CUHK. We thank Prof. S. H. Wei for helpful discussions.

* Corresponding author.

jyzhu@phy.cuhk.edu.hk

# Supplemental Information: Step Stone Effect: An *sp* anti-bonding Orbital Mediated Long-Range Ferromagnetism in Cr-doped Carrier-Free Bi$_2$Te$_3$

Chunkai Chan, Xiaodong Zhang, Yiou Zhang, Kinfai Tse, Bei Deng, Jingzhao Zhang and Junyi Zhu[*]

Department of Physics, Chinese University of Hong Kong, Hong Kong

**I. Computational methods and details**

All calculations are performed using projected augmented wave (PAW) [S1] potentials with Perdew-Burke-Ernzerhof (PBE) [S2] generalized gradient approximation (GGA) as implemented in Vienna *ab initio* simulation package (VASP) [S3]. The cutoff energy for plane-wave expansion was set to 350eV for both Bi$_2$Se$_3$ and Bi$_2$Te$_3$. Gamma centered 3×3×1 *k* mesh is used to sample the Brillouin zone for 4×4 supercell of Bi$_2$X$_3$ (X=Se, Te), as shown in Fig. S1, with all the neighboring sites marked. The number 0 represents the position of first Cr atom. The numbers 1 to 7 represent the 1st to 7th nearest neighbor placements of the second Cr atom. All atoms in every supercell are fully relaxed until the residual force is less than 0.01 eV/Å. Convergence tests about kpoints, cell sizes, vacuum size, magnetism, and energy cutoffs have been performed. The formation energy of the dopants is defined as:

$$\Delta H_f(Cr) = E_{tot}(Bi_2X_3:Cr) - E_{tot}(host) - \Sigma_i n_i \mu_i \quad (1),$$

where $E_{tot}$ (Bi$_2$X$_3$:Cr) (X=Se, Te) is the total energy of a supercell with Cr dopants; $E_{tot}$ (host) is the total energy of the supercell without impurities; $n_i$ is the number of certain atoms added to ($n_i$ < 0) or removed from ($n_i$ > 0) the supercell; and $\mu_i$ is the corresponding chemical potential. In the main text, we take the relative formation energy in reference to the lowest energy among all configurations.

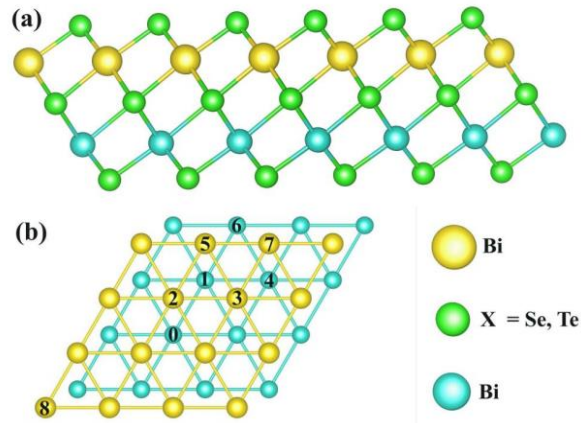

Fig. S1. (Color online) (a) The side view of one (4 × 4) quintuple layer $Bi_2Se(Te)_3$ supercell with green atoms being Se(Te), yellow and blue atoms being the Bi atoms in the upper and lower layers respectively. (b) The top view of the upper (yellow) and lower (blue) layer of Bi atoms.

## II. Spin orbital coupling (SOC) effect

The strong SOC in $Bi_2Se_3$ and $Bi_2Te_3$ cause band inversion between valence band and conduction band, and results in different occupations of electrons [S4], which may change the long-range magnetic order. Besides, the strong electron-electron correlation for $d$-electrons of Cr atoms requires introduction of $U$ values to account for the on-site Columbic interaction. Previously it was found that inclusion of the electron-electron correlation enlarges the band gap and slightly changes the electron occupations [S5].

To investigate the effects of SOC and +U on the long-range magnetic order, we performed GGA + U [S6] calculations on Cr-doped $Bi_2X_3$ (X=Se, Te) with SOC effect included. For one quintuple layer of $Bi_2X_3$ (X=Se,Te), a 4×4 unit-cell slab and at least 20 Å vacuum are included, and 3x3x1 $k$-point sampling mesh is adopted. $U$ = 3eV and $J$ = 0.87eV are applied on Cr $d$-orbitals, same values as those used in previous work [S5]. The two Cr atoms in the supercell are put either in the second nearest neighbor (2nd-NN) or in the seventh nearest neighbor (7th-NN). Results are summarized in Fig. S2.

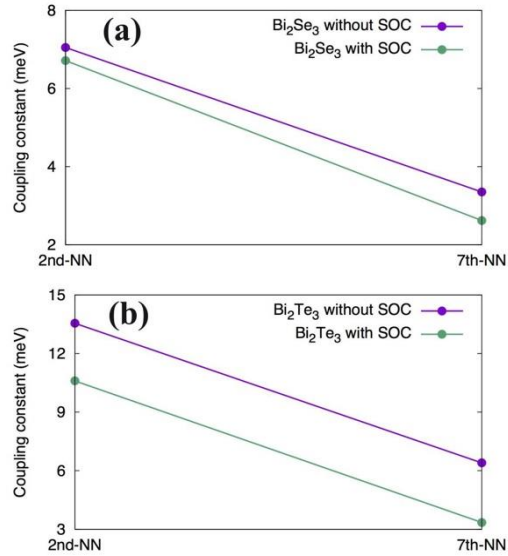

Fig. S2. Ferromagnetic coupling constant (defined by half of energy difference between AFM state and FM state) of (a) Cr doped $Bi_2Se_3$ and (b) Cr doped $Bi_2Te_3$.

Compared with calculations without SOC effect and $+U$ effect, the ferromagnetic coupling strength in both systems becomes weaker, as $+U$ treatment usually leads to more localized states for $d$-orbitals. Nevertheless, magnetic coupling at the 7th-NN still preserves. Although the SOC changes the positions of the *p* orbitals of Bi and anions, the *sp* anti-bonding state still mediates the magnetic interactions.

In addition, we calculated the formation energy of Cr-doped $Bi_2X_3$ (X=Se, Te) and included the SOC. The FM configuration at 7th-NN is still the most stable, about 2meV lower than the second lowest energy configuration. However, the most stable configuration of $Bi_2Se_3$ is still the 2nd-NN, about 3meV lower than the 7th-NN.

**III. Partial charge of *sp* hybridized orbital**

The *sp* hybridization also can be visualized by partial charge density as shown in Fig. S3 (partial charge density of the second band below VBM). The *sp* hybridized orbital around Bi is clearly shown. This hybridized orbital is the step stone state for long range magnetic interaction in $Bi_2Te_3$.

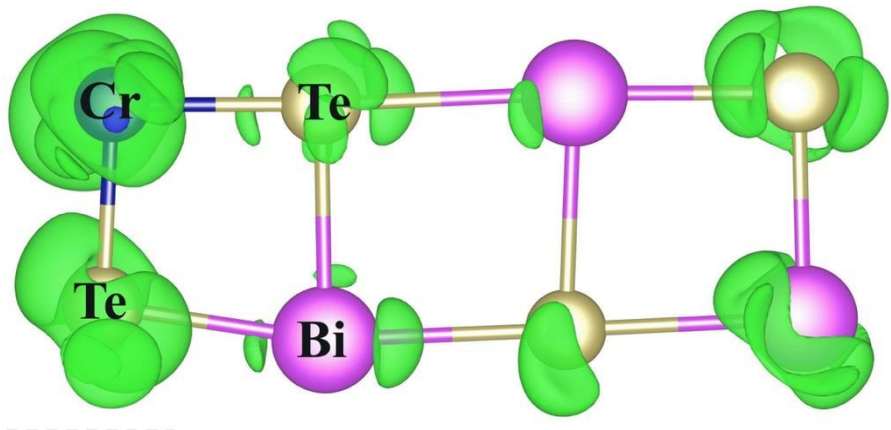

Fig. S3 Partial charge density of the second band below VBM

*Corresponding author

jyzhu@phys.cuhk.edu.hk